\documentstyle[prb,aps,multicol,graphics]{revtex}

\begin{document}
\title{Weak localization in the 2D metallic regime of Si-MOS}
\author{G.~Brunthaler$^{1}$, A.~Prinz$^{1}$, G.~Bauer$^{1}$,
V.\,M.~Pudalov$^{1,2}$, E.\,M.~Dizhur$^3$\\ %
J.~Jaroszynski$^{4}$, P.~Glod$^{4}$ and T.~Dietl$^{4}$}
\address{$^{1}$ Institut f\"{u}r Halbleiterphysik, Johannes Kepler
Universit\"{a}t, A-4040 Linz, Austria\\ %
$^{2} $P.\ N.\ Lebedev Physics Institute of the Russian Academy of
Sciences, Moscow 117924, Russia \\ %
$^3$High Pressure Physics Institute, Troitsk, Moscow district
142092, Russia \\ %
$^4$Institute of Physics, Polish Academy of Sciences, PL-02668
Warszawa, Poland}
\maketitle
\centerline{{\tt e-mail:G.Brunthaler@hlphys.uni-linz.ac.at}}

\begin{abstract}
The negative magnetoresistance due to weak localization is
investigated in the two-dimensional metallic state of Si-MOS
structures for high conductance values between 35 and 120 $e^2/h$.
The extracted phase coherence time is equal to the momentum
relaxation time at 10\,K but nearly 100 times longer at the lowest
temperature. Nevertheless, only weak logarithmic corrections to
the conductivity are present in the investigated temperature and
concentration range thus proving the absence of strong quantum
effects due to electron-electron interaction. From saturation
effects of the phase coherence time a lower boundary for
spin-orbit scattering of about 200\,ps is estimated.
\end{abstract}

\begin{multicols}{2}
\section{Introduction}
Strong evidence for a metal-insulator transition (MIT) was
observed in Si metal-oxid-semi\-con\-duc\-tor (MOS) structures
\cite{krav94,Popo97} and other two-dimensional (2D) semiconductor
systems \cite{experiments}. One of the most striking features of
this phenomenon is a strong exponential drop of the resistivity
$\rho = \rho_0 + \rho_1 \exp(-(T_0/T)^q)$ which saturates at low
temperatures. According to scaling arguments for non-interacting
2D electron systems no such MIT was expected \cite{Abra79}.
However, for interacting 2D systems, theoretical evaluation showed
that the MIT in 2D is not forbidden by basic arguments
\cite{positive-beta}. %
Now the central question about the MIT in 2D concerns its origin.
If electron-electron interaction is responsible for the strong
decrease in resistivity, then quantum effects should dominate the
metallic regime \cite{qu-effects}. If, on the other hand, the
scattering mechanism across a spin gap \cite{Puda97} or the sharp
crossover from quantum to classical transport \cite{DasSarma99}
takes place, then the transport properties should not be dominated
by quantum effects but should be rather explainable by the
classical Boltzmann transport behavior.

We have thus investigated the negative magnetoresistance due to
the weak-locali\-zation effect \cite{WL-theory} in order to get
information about the phase coherence time $\tau_\varphi$ in a
regime where possibly electron-electron interactions cause the
anomalous metallic state. For carrier densities around the
critical concentration $n_c$, conductivity corrections from single
electron backscattering and interaction effects in the
electron-electron and electron-hole channels overlap
\cite{interaction} and it is not possible to get an unambiguous
value for the phase coherence time. Therefore our investigations
were focused on the high carrier density regime with conductance
$g = \sigma /(e^2/h)$ between 35 and 120, where the strong
exponential drop in resistivity has shifted to higher temperatures
and weak $\ln(T)$ terms with both, negative and positive sign
dominate the behavior \cite{Puda98}. In this regime, the negative
magnetoresistance due to weak localization is restricted to a
narrow range of magnetic fields and can be evaluated under the
assumption that the interaction terms have only small influence on
the negative magnetoresistance.

\section{Measurements and discussion}

Our investigations were performed on two high-mobility Si-MOS
samples Si-43 and Si-15 with peak mobilities of $\mu = 20,000$ and
32,000\,cm$^2$/Vs, respectively. The samples consist of 5\,mm long
and 0.8\,mm wide Hall bars, covered with a 200\,nm thick Si-Oxide
layer serving as an insulator and a 1000\,nm thick Al-gate on top.
The resistivity and Hall measurements were performed in a four
terminal ac-technique at a frequency of 19\,Hz.

The magnetoresistivity measurements were performed on sample Si-43
at electron densities $n$ from $5.4 \times 10^{11}$ to $3.5 \times
10^{12}$\,cm$^{-2}$ and temperatures $T$ between 0.29 and 10\,K
and on sample Si-15 for $1.05 \times 10^{12} < n < 4.5 \times
10^{12}$\,cm$^{-2}$ and $0.3 < T < 1.4$\,K. In this density range,
the conductance $g$ is between 35 and 120, which is just the
region below the maximum metallic conductivity in these Si-MOS
structures \cite{Puda99}. The open circles in Fig.~1 represent the
negative magnetoresistance $\Delta\rho(B) = \rho_{xx}(B) -
\rho_{xx}(0)$ at a density of $n = 1.05 \times 10^{12}$\,cm$^{-2}$
at different temperatures for Si-15. The negative
magnetoresistance was fitted to the single-electron weak
localization correction to conductivity \cite{WL-theory}
\begin{equation}
\Delta\sigma_{xx} = - \frac{\alpha g_\nu e^2}{2\pi^2\hbar}
 \left[\Psi\left(\frac{1}{2}+\frac{\hbar}{4eBD\tau}\right)
       -\Psi\left(\frac{1}{2}+\frac{\hbar}{4eBD\tau_{\varphi}}\right)\right],
\label{eq:WL}
\end{equation}
where $\Psi$ is the Digamma function, $B$ is the applied
perpendicular magnetic field, $D$ the diffusion coefficient and
$\tau$ the momentum relaxation time. The values for $D$ and $\tau$
were deduced from the temperature and density dependent Hall and
resistivity measurements. The prefactor $g_{\nu}$ describes the
valley degeneracy and $\alpha$ depends, according to theory
\cite{Fuku81}, on the ratio of intra-valley to inter-valley
scattering rates and should be between 0.5 and 1. We found values
between 0.5 and 0.6 for $\alpha$, which is in the expected range.
The full lines in Fig.~1 represent the best least square fits
through the data points according to Eq.~1 for the different
temperatures.

\begin{figure}
\begin{center}
\resizebox{0.80\linewidth}{!}{\includegraphics{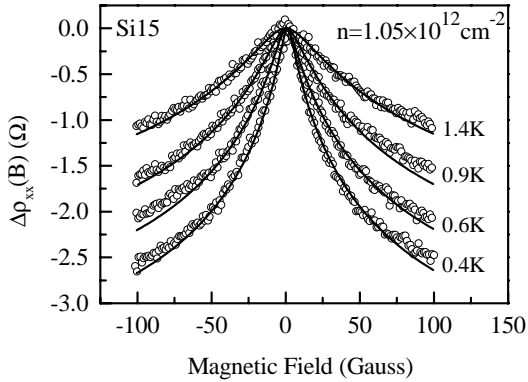}}
\begin{minipage}{8.2cm}
\vspace{2mm} %
\caption{Change of resistivity $\Delta\rho(B) =
\rho_{xx}(B) - \rho_{xx}(0)$ versus magnetic field $B$ at an
electron density of $1.05 \times 10^{12}$\,cm$^{-2}$ for sample
Si-15 at temperatures of 0.4, 0.6, 0.9 and 1.4\,K. The open
circles represent the measurements and the full line the best
least square fit of the data according to the weak localization
dependence of (1).} \label{fig:rho(B)}
\end{minipage}
\end{center}
\end{figure}

The influence of stray magnetic fields was tested by carefully
shielding the sample together with a copper coil by a $\mu$-metal
foil for temperatures down to 100\,mK. As no influence was
observed, we conclude that the width and height of the weak
localization peak is not disturbed by any background magnetic
fields. We also have not seen any sign for a superconducting state
of the 2D electron gas which might be disturbed by very small
magnetic fields.

In Fig.~2, the temperature dependence of the phase coherence time
$\tau_\varphi$ is shown for several carrier densities in the range
from $5.4 \times 10^{11}$ to $3.5 \times 10^{12}$\,cm$^{-2}$ for
Si-43. It is found that $\tau_\varphi$ increases by about a factor
of 100 from 1\,ps at 10\,K to nearly 100\,ps at 0.29\,K. Similar
values for $\tau_\varphi$ were also reported for earlier
investigations on Si-MOS structures \cite{old-WL-exp}. The
increase of $\tau_\varphi$ can be described approximately by a
power law $\tau_\varphi \propto T^{-p}$. At higher temperatures,
large momentum transfer processes dominate the electron-electron
scattering and $p = 2$ is expected, whereas at low temperatures,
small momentum transfer dominates in disordered systems even for
$k_F \ell \gg 1$ and cause $p = 1$ \cite{interaction}.

According to the data in Fig.~2, $p$ is decreasing slightly from
1.4 above 4\,K to 1.3 below 1\,K for the lowest density, whereas
$p$ shows a strong change from 1.5 towards 0.5 in the same
temperature interval for the highest density. The latter value of
0.5 can not be explained by theory and the strong change in the
slope of Fig.~2 indicates rather a saturation of the phase
coherence time below 0.3\,K. Such a saturation of $\tau_\varphi$
can be caused by electron heating effects due to high frequency
noise or by other processes which limit the height of the weak
localization effect. According to Hikami et al.~\cite{Hika80},
such a limitation can be caused by spin scattering due to magnetic
impurities or spin-orbit interaction. From the additional terms
for spin scattering and spin-orbit scattering, which enter (1),
such a limiting scattering time can be estimated to be around
200\,ps. But from the available data no definite conclusion about
the process, which limits $\tau_\varphi$, can be drawn.

In the temperature range where $\tau_\varphi$ increases by about a
factor 100, the resistivity of the sample is nearly constant.
Especially for the highest density of $n = 3.5 \times
10^{12}$\,cm$^{-2}$, the exponentially strong changes in the
conductivity, which where observed below 1\,K for densities near
the MIT, have shifted to temperatures above 10\,K. Only weak
changes $\Delta \rho \propto \ln(T)$ remain at small $T$
\cite{Puda98,Puda99} despite the strong increase of
$\tau_\varphi$.

\begin{figure}
\begin{center}
\resizebox{0.80\linewidth}{!}{\includegraphics{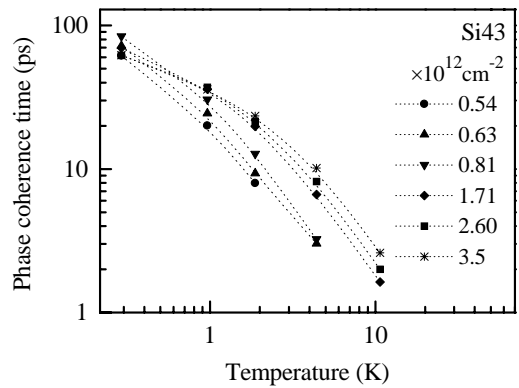}}
\begin{minipage}{8.2cm}
\vspace{2mm} %
\caption{Temperature dependence of the phase coherence time
$\tau_\varphi$ for densities of 0.54, 0.63, 0.81 1.71, 2.6 and
$3.5 \times 10^{12}$\,cm$^{-2}$ for sample Si-43. The dashed lines
are guides for the eye.} \label{fig:tphi(T)}
\end{minipage}
\end{center}
\end{figure}

The dependence of $\tau_\varphi$ on the carrier concentration is
depicted in Fig.~3. For temperatures in the range from 1 to
10.7\,K, $\tau_\varphi$ increases with increasing carrier
concentration. Below $n = 1 \times 10^{12}$\,cm$^{-2}$, this
increase is steep, whereas for higher densities the increase
becomes smaller. At a temperature of 0.29\,K, $\tau_\varphi$
increases as well below $n = 1 \times 10^{12}$\,cm$^{-2}$, but
decreases slightly for higher concentrations. This abnormal
behavior at 0.29\,K can be assigned to the observed saturation of
$\tau_\varphi$ in the temperature dependence for higher densities
as described above. The same investigations for sample Si-15
showed a very similar behavior of the phase coherence time
$\tau_\varphi$ with saturation effects near $T = 0.3\,K$.

In Fig.~3 the momentum relaxation time $\tau$ is shown for
comparison. It was calculated from the conductivity $\sigma$ at $B
= 0$ as $\tau = m^* \sigma / n e^2$ with $m^* = 0.19 m_0$ the
transversal effective mass and $n$ the carrier density as derived
from the Hall coefficient $R_H$. As the magnetoresisitivity due to
weak localization is a very small effect ($\Delta \sigma / \sigma
\approx 10^{-3}$), there is practically no difference whether
$\tau$ is calculated at $B = 0$ or at $B > \Phi_0 /
\ell_\varphi^2$. From Fig.~3, it can be seen that the overall
behavior of $\tau_\varphi$ and $\tau$ is quite different and
independent of each other. This can be understood by the different
underlying scattering mechanisms for inelastic and elastic
processes. The elastic processes are due to impurity and surface
roughness scattering whereas the inelastic processes are caused by
electron-phonon and electron-electron scattering.

\begin{figure}
\begin{center}
\resizebox{0.80\linewidth}{!}{\includegraphics{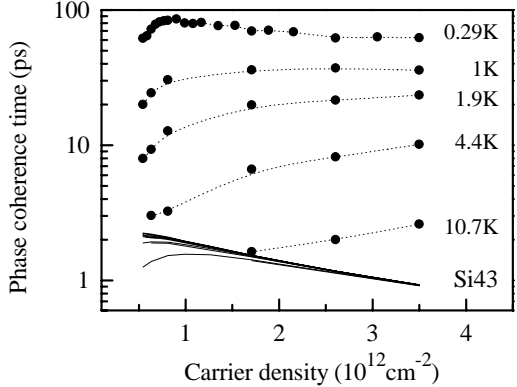}}
\begin{minipage}{8.2cm}
\vspace{2mm} %
\caption{Electron density dependence of the phase coherence time
$\tau_\varphi$ for temperatures of 0.29, 1.0, 1.9, 4.4 and 10.7\,K
(symbols, dashed lines are guide for the eye) for sample Si-43.
For comparison, the momentum relaxation time $\tau$ is shown by
full lines where lower lying lines at low densities represent
higher temperatures.} \label{fig:tphi(n)}
\end{minipage}
\end{center}
\end{figure}

It should be noted that we observe a negative magnetoresistance
within $\pm 50$ Gauss despite the fact that there is a
superconducting Al-gate on top of the Si-MOS structures. Bulk Al
is a type I superconductor with a critical superconducting
temperature of $T_c = 1.175$\,K and a critical magnetic field of
$H_c = 105$\,G. We have tested the superconductivity of our
Al-gate directly by performing a fore-wire resistance measurement.
We found a sudden decrease of the resistivity to zero at a
temperature of 1.20\,K. At 0.29\,K, the zero-resistivity could be
quenched by an external magnetic field of 51\,G. The deviation of
this field from $H_c$ is caused by the intermediate state which
has a finite resitivity when the superconducting regions are not
connected any more \cite{intermediate}. In addition, the
superconducting properties may be modified as the Al-gate consists
of a thin evaporated layer which is expected to be strongly
disordered.

The effect of a superconducting gate on the weak localization was
treated in detail by Rammer and Shelankov \cite{Rammer87}. At
magnetic fields below $H_c$, the magnetic flux is collected in the
flux tubes of the intermediate state. At the position of the 2D
electron gas, 200\,nm away from the superconducting gate, the
nonuniformity of the magnetic field persists only if the period of
the magnetic structure is larger than this distance. The
characteristic lateral size of the domains in the intermediate
state can be estimated from the laminar model \cite{intermediate}.
For pure Al, one gets a period $a$ of about 300\,nm, whereas for a
strongly disordered material a period of rather 80\,nm is
expected. This domain period has to be compared with the phase
coherence length $\ell_\varphi = \sqrt{D \tau_\varphi}$, with $D$
the diffusion coefficient. From our investigations it follows that
$\ell_\varphi$ is in the range from 600 to 1000\,nm at
temperatures below $T_c = 1.2$\,K. As this range of values is much
larger than the typical domain period $a$, no influence of the
superconducting gate on the negative magnetoresistivity is
expected \cite{Rammer87}. This is in agreement with our
measurements, where we have observed no direct influence.

\section{Conclusions}
We have investigated the weak localization in the metallic regime
of Si-MOS structures at high carrier density where the conductance
is between 35 and 120 $e^2/h$. When decreasing the temperature
from 10 to 0.29\,K, the phase coherence time $\tau_\varphi$
increases from 2 to about 100\,ps enabling in principle strong
quantum effects to take place. Nevertheless, only weak $\ln(T)$
type changes of the conductivity were observed from 10\,K where
the phase coherence is as short as the momentum relaxation time
down to 0.29\,K where $\tau_\varphi$ is about 100 times larger. We
thus conclude that in the investigated high conductance regime no
strong quantum effects due to electron-electron interaction take
place which could drive the system into the metallic state. From
saturation effects of $\tau_\varphi$, we can estimate a lower
boundary for the spin-orbit interaction scattering time of about
200\,ps.

The work was supported by RFBR, by the Programs on "Physics of
solid-state nanostructures" and "Statistical physics", by INTAS,
by NWO, by the "Fonds zur F\"{o}rderung der Wissenschaften"
P-13439 and GME Austria, and by KBN (Grant No 2-P03B-11914)
Poland.

\end{multicols}

\end{document}